\documentclass[aps,showpacs,nofootinbib,pra,superscriptaddress,twocolumn,floatfix]{revtex4-1}
\usepackage{multirow}
\usepackage{amsfonts}
\usepackage{amsmath,mathtools}
\usepackage{amssymb}
\usepackage{graphicx}
\usepackage{dcolumn}
\usepackage{times}
\usepackage{color}
\usepackage{ulem}
\usepackage{hhline}

\usepackage{alphabeta}
 \usepackage[LGR, T1]{fontenc}
 \usepackage{graphicx}
 \usepackage{tikz}
\usepackage{anyfontsize}
\usepackage{caption}
\usepackage{makecell}
\usepackage{array}
\newcolumntype{?}{!{\vrule width 1pt}}
\usepackage{slashed}

\usepackage{amssymb}
 \usepackage{cancel}
 \usepackage{dsfont}
 \usepackage{xcolor}

\usepackage{empheq}
\usepackage{extarrows} 
\usepackage[mathscr]{euscript}
 \usepackage{fullpage}
\newcolumntype{C}[1]{>{\centering\let\newline\\\arraybackslash\hspace{0pt}}m{#1}}
  \usepackage[hidelinks]{hyperref}
  
\begin{document}

 \title{Generalized Continuity Equations  for  Schr\"odinger and 
 Dirac Equations }
\author{A. Katsaris}
\address{Department of Physics, University of Athens, GR-15784 Athens, Greece}

\author{P.~A. Kalozoumis}
\address{Department of Engineering and Informatics, Hellenic American University, 436 Amherst Street, Nashua, NH 03063, USA}
\address{Materials Science Department, School of Natural Sciences, University of Patras, Patras 265 04, Greece}

\author{F.~K.~Diakonos}
\email[]{fdiakono@phys.uoa.gr}
\address{Department of Physics, University of Athens, GR-15784 Athens, Greece}

\date{\today}

\begin{abstract}
The concept of the generalized continuity equation (GCE) was recently introduced in [J. Phys. A: Math. and Theor. {\bf 52}, 1552034 (2019)], and was derived in the context of $N$ independent Schr\"{o}dinger systems.
The GCE is induced by a symmetry transformation which mixes the states of these systems, even though the $N$-system Lagrangian does not. As the $N$-system Schr\"{o}dinger Lagrangian is not invariant under such a transformation, the GCE will involve source terms which, under certain conditions vanish and lead to conserved currents. 
These conditions may hold globally or locally in a finite domain, leading to globally or locally conserved currents, respectively. 
In this work, we extend this idea to the case of arbitrary $SU(N)$-transformations and we show that a similar GCE emerges for $N$ systems in the Dirac dynamics framework. The emerging GCEs and the conditions which lead to the attendant conservation laws provide a rich phenomenology and potential use for the preparation and control of fermionic states.
\end{abstract}

\maketitle
 \section{Introduction}
 
Symmetries and the emerging conservation laws are of fundamental importance in Physics. For continuous transformations, the link between conservation laws and symmetries is provided by Noether's theorem~\cite{Noether1918}. Based on the Lagrangian formulation of the underlying dynamics, Noether's theorem allows for applications both in classical field theory and in non-relativistic quantum mechanics, leading to symmetry induced continuity equations directly linked to conservation laws. In the context of Schr\"odinger dynamics, these continuity equations usually refer to a single quantum state of the considered system and they are connected to internal symmetries of the Hilbert space.
 
Recently, it was shown that symmetry-induced bilocal continuity equations emerge in quantum Hermitian and non-Hermitian systems~\cite{Spourdalakis2016} from symmetry transformations leaving the corresponding Lagrangian density invariant. They generalize the usual continuity equations, involving two distinct wave fields which obey different dynamics, when the non-Hermitian part in the potential term is non-zero. The corresponding two-field Lagrangian is invariant under dilatation and phase transformations of these fields. The theoretical framework developed in \cite{Spourdalakis2016} addresses the impact of \textit{global} symmetry transformations on quantum Schr\"odinger states and indicates that the currents associated with the bilocal continuity equations act as correlators between the two fields. However, the existence of global symmetries often can be an idealized scenario. In realistic systems symmetries can be broken due to defects, impurities and boundary conditions. Moreover, the presence of a potential term depending only on space leads to stationary states, which violate the equal treatment between space and time, indicating the breaking of Lorentz invariance.

A case of particular interest is when a symmetry transformation, even though globally broken, appears to hold in restricted spatial domains. The existence of such \textit{local} symmetries, being present to a broad range of photonic~\cite{Kalozoumis2013b}, acoustic~\cite{Kalozoumis2015a} and, quantum~\cite{Kalozoumis2013a} systems induce interesting spectral properties and exhibit a new class of non-local currents (NLC) which have been shown to characterize rigorously the symmetry breaking~\cite{Kalozoumis2014a}. The question is, if these currents, which are connected to local symmetries, can be extracted variationally from a suitable Lagrangian density. Clearly, the GCE derived in~\cite{Spourdalakis2016} is not directly transferable to a Lagrangian density possessing local symmetries, and a more general mathematical framework is needed for the description of local conservation laws assigned to systems with local symmetries. Such an approach has been recently proposed in~\cite{Diakonos2019}. The variational scheme developed there, utilizing the concept of a super-Lagrangian, allowed to derive a GCE obeyed by the local symmetry induced NLC. This mathematical framework led to GCEs linking pairwise wave fields which belong to different Schr\"odinger problems and involve the dynamics of single particle evolution in different potentials $V_{i}(\mathbf{x})$ with $i=1,...,N$. Within this scenario, the extended symmetry transformation applied to the super-Lagrangian is chosen to belong to the  $SU(N)$ Lie group and its algebraic form determines the form of the occurring source terms. As it has been shown~\cite{Diakonos2019}, under certain conditions the source terms vanish and GCEs lead to generalized conservation laws. This consistent variational framework, allowed the description of a special class of currents, which are locally conserved in restricted spatial domains whenever symmetries of the potential terms are valid within these domains. From a group theoretical point of view, this analysis was restricted to transformations, which involved only generators of the $SU(N)$ Cartan sub-algebra and the super-Lagrangian was designed to describe only Schr\"odinger dynamics.

Although, the Schr\"odinger equation can handle a multitude of quantum mechanical problems, it is not suitable to describe microscopic systems at high energies, such as relativistic fermions. The need to incorporate special theory of relativity in a quantum mechanical context led to the Dirac equation, formulated by Paul Dirac in 1928. In this work, we extend the concept of the super-Lagrangian~\cite{Diakonos2019} for non-relativistic Schr\"odinger problems, to a relativistic framework. We introduce GCEs which contain Dirac fermionic states and we consider $SU(N)$ symmetry transformations with generators beyond the Cartan sub-algebra, covering this way all the possible $SU(N)$ transformations. Thus, the set of possible source terms in the GCEs will be significantly increased, giving rise to a novel class of symmetry-induced invariant currents. Our results provide a unified treatment of generalized conservation laws in the presence of potentials with globally broken, domain-wise sustained symmetries for both Dirac and Schr\"odinger-type problems.

The paper is organized as follows: In Section~\ref{sec_II} we introduce the super-Lagrangian for Dirac fermions, we derive the corresponding GCE, and extract the global and local conservation laws that emerge. In Section~\ref{sec_III} we apply the concept of the super-Lagrangian to the case of two, single-Fermion Dirac systems and we derive the corresponding global and local conservation laws. In Section~\ref{sec_IV}, non-relativistic systems are considered again and we generalize the results found in~\cite{Diakonos2019} for $SU(N)$ symmetry transformations. Our results are summarized in Section~\ref{sec_V}. 

\section{Super-Lagrangian for Dirac Fermions and the Generalized Continuity Equation}\label{sec_II}

The Dirac Lagrangian density $\mathcal{L}_i$ is,
\begin{equation}
\mathcal{L}_i=\bar\psi_i(i\slashed{\partial}-V_i(\vec x))\psi_i
\label{eq:1}
\end{equation}
where $\psi_i$ is a Dirac spinor. $\mathcal{L}_i$ describes the dynamics of fermionic states in the presence of the potential term $V_i(\vec x)$. Applying the usual variational approach to $\mathcal{L}_i$ leads to the Dirac equation for the field $\psi_i$. The super-Lagrangian is defined as the sum of the individual Lagrangian densities which describe the dynamics of independent quantum systems in different potential landscapes $V_i(\vec x)$, $i=1,\dots,N$. For these $N$ independent Dirac systems the super-Lagrangian can be written as, 
 \begin{equation}
\label{eq:2}
\mathcal{L}=\sum_{i=1}^N\mathcal{L}_i=\bar\Psi(H_0-V)\Psi=\mathcal{L}_0+\tilde{\mathcal{L}}
\end{equation}
 where $\Psi$, $\bar\Psi$ are defined as, 
\begin{equation}
\label{eq:3}
\Psi=\begin{pmatrix} \psi_1 \\ \psi_2 \\ \cdots \\ \psi_N    \end{pmatrix}
\end{equation}
and
\begin{equation}
\label{eq:4}
 \bar\Psi= \Psi^\dagger\otimes\gamma^0=\left(\begin{array}{cccc} \bar{\psi}_1 & \bar{\psi}_2 & \cdots & \bar{\psi}_N \end{array}\right)
\end{equation}
respectively and $\psi_i$ are the corresponding Dirac spinors.

The quantities $\mathcal{L}_0$ and $\tilde{\mathcal{L}}$ are defined as, 
\begin{equation}
\mathcal{L}_{0}=\bar\Psi H_0\Psi ~~~;~~~ \tilde{\mathcal{L}}=-\bar\Psi V\Psi
\label{eq:5}
\end{equation}
with
\begin{equation}
\label{eq:6}
H_0=i\slashed{\partial}\otimes\mathds{1}_N
\end{equation}
and
\begin{equation}
\label{eq:7}
V=\text{diag}(V_1,V_2,\cdots,V_N),
\end{equation}
being the kinetic and potential terms, respectively.

To derive the GCEs from the super-Lagrangian in Eq.~(\ref{eq:2}), we make an arbitrary transformation of the spinors $\Psi$. Note, that the source terms in the resulting GCEs are subject to this transformation. 
We are interested in the conditions under which these source terms vanish. In this case, we consider an $SU(N)$ transformation of $\Psi$, which for an infinitesimal variation $\delta \theta_a$ can be expressed as 
\begin{equation}
\label{eq:8}
U \Psi =e^{-i\delta\theta_aT^a} \Psi \simeq(1-i\delta\theta_a T^a) \Psi
\end{equation}
where $T_a$ denotes the generators of the $SU(N)$ group and $\delta\theta_a$ are arbitrary small parameters. 
The variation of the Lagrangian density $\mathcal{L}$ yields that $\delta \mathcal{L}_0=0$,  $\delta \mathcal{\tilde L}\neq0$ and consequently $\delta \mathcal{L}=\delta\mathcal{\tilde L}\neq0$, with $\delta \mathcal{\tilde L}=\mathcal{\tilde L}'-\mathcal{\tilde L}$. These considerations lead to a current conservation law with a vanishing $4-$divergence.

If $\mathcal{L}$ was invariant under the transformation $U$ 
 then, according to Noether's theorem, a typical continuity equation would be obtained, associated with this invariance. However, here $\mathcal{L}$ is not invariant, since $\delta\mathcal{\tilde L} \neq 0$, and Noether's approach leads to a GCE with non-vanishing source terms, dictated by the performed symmetry transform in Eq.~\eqref{eq:8}.  
In order to calculate $\delta \mathcal{\tilde L}$ we employ the generators $T_a$ of the $SU(N)$ group and we write the potential matrix $V$ as,
\begin{equation}
\label{eq:9}
V(x)=V_0(x)\mathds{1}_N+\sum_{k=1}^{N^2-1} C_k(x)T_k
\end{equation}
where
\begin{equation*}
\
V_0(x)=\frac{Tr(V\cdot \mathds{1}_N)}{N} ~~~ ; ~~~ C_k(x)=Tr(V\cdot T_k).
\end{equation*}
The explicit form of $C_{k}(x)$ and the details for their derivation are presented in Appendix~\ref{Vapp}.
The fact that $\delta \mathcal{L}=\mathcal{\tilde L}'-\mathcal{\tilde L}$ only, leads to
\begin{equation}
\label{eq:10}
\delta\mathcal{L}=- \bar\Psi(U^\dagger V U - V)\Psi. 
\end{equation}
Using Eq.~\eqref{eq:9} it is straightforward to find that
\begin{equation}
\label{eq:11}
U^\dagger V U - V=-\delta\theta_a C_kf_{ak}^cT_c
\end{equation} 
and consequently
\begin{equation}
\label{eq:12}
\delta \mathcal{L}=\delta\theta^a f_{abc}\bar\Psi C_bT_c\Psi.
\end{equation}

On the other hand the variation $\delta\mathcal{L}$ of the Lagrangian can be also written as,
\begin{align}
\label{eq:13}
\delta\mathcal{L}=\sum_{i=1}^N [ \frac{\partial \mathcal{L}}{\partial \psi_i}\delta\psi_i+\frac{\partial \mathcal{L}}{\partial( \partial_\mu\psi_i)}\delta(\partial_\mu\psi_i) \nonumber \\+\frac{\partial \mathcal{L}}{\partial \bar{\psi_i}}\delta\bar{\psi_i}+\frac{\partial \mathcal{L}}{\partial (\partial_\mu{\bar{\psi_i})}}\delta(\partial_\mu\bar{\psi_i})]
\end{align}
where $i$ enumerates the Dirac spinors. Employing the Euler-Lagrange equations for each fermion $i$,
\begin{align}
\label{eq:14}
\frac{\partial \mathcal{L}}{\partial \psi_i}-\partial_\mu\big({\frac{\partial \mathcal{L}}{\partial( \partial_\mu\psi_i)}}\big)=0  \nonumber \\
\displaystyle{\frac{\partial \mathcal{L}}{\partial \bar{\psi}_i}}-\partial_\mu\big(\displaystyle{\frac{\partial \mathcal{L}}{\partial (\partial_\mu\bar{\psi}_i)}}\big)=0
\end{align}
and the relations,
\begin{equation}
\label{eq:15}
\frac{\partial \mathcal{L}}{\partial (\partial_\mu\bar\psi_i)}=0 ~~,~~
\delta(\partial_\mu\psi_i)=\partial_\mu(\delta\psi_i).
\end{equation}
we find that the Lagrangian variation is given by
\begin{equation}
\label{eq:16}
\delta\mathcal{L}= \sum_{i=1}^N [\partial_\mu\big({\frac{\partial \mathcal{L}}{\partial \partial_\mu\psi_i}}\delta\psi_i\big)].
\end{equation}
Using Eq.~\eqref{eq:9} and after some algebra, we find 
\begin{equation}
\label{eq:17}
\delta\mathcal{L}=\partial_\mu(\bar{\Psi}(\gamma^\mu\otimes\mathds{1}_N)T_a\Psi )\delta \theta^a
\end{equation} 		
Combining Eq.~\eqref{eq:12} and Eq.~\eqref{eq:17} and keeping only first order terms in $\delta\theta^a$ we obtain a GCE for the Dirac equation,
 \begin{equation}
 \label{eq:18}
\partial_\mu(\bar{\Psi}\left(\gamma^\mu\otimes T_a\right)\Psi )=f_{abc}\bar\Psi \left(C_b\otimes T_c\right)\Psi,
\end{equation}
which is associated to $SU(N)$ symmetry transformations. Equation~\eqref{eq:18}
is one of the core results of this work. The term ``generalized'' in the derived continuity equation has a two-fold meaning. Firstly, Eq.~(\ref{eq:18}) involves non-zero source terms and therefore it does not, directly, lead to a 4-current conservation. Secondly, the 4-current $J^\mu_a=\bar{\Psi}\gamma^\mu T_a\Psi$ is not the usual current, but an abstract mathematical quantity that connects the solutions of \textit{different} Dirac problems [see the definition of $\Psi$ in Eq.~\eqref{eq:3}]. Thus, Eq.~\eqref{eq:18}, in fact, corresponds to $N^2-1$ different equations and the number of the emerging 4-currents is equal to the number of the $SU(N)$ group's generators. 
In the case of a diagonal potential matrix $V$, as the one discussed here, and due to the property $f_{aac}=0$ of the structure constants $f_{abc}$, the remaining not trivial currents will be those related to the non-diagonal elements of an $SU(N)$ matrix in the fundamental representation. These currents are not all independent of each another. They appear in Hermitian conjugate pairs and, in this sense, for every emerging current, there is a corresponding Hermitian conjugate, which is described by Eq.~\eqref{eq:18}, as well.
 
Of particular interest is the case where the source terms of the GCE in Eq.~\eqref{eq:18} vanish, leading to a new class of conservation laws. Here we will examine the possible scenarios which lead to vanishing source terms in Eq.~\eqref{eq:18}. These can be classified in the following two categories:
(i) {\it Global conservation laws}. The source terms become zero due to a global property of the potential and holds for any spinor $\Psi$. In this case the emerging conservation law will be valid globally. 
(ii) {\it Local conservation laws}. The source terms become zero due to a local property of the potential and holds for any $\Psi$.
As local we consider here a property of the potentials which is valid within a finite spatial domain. The emerging conservation law will also be valid only within the spatial region where this property of the potentials holds.

In principle, there is a third category related to the accidental case where the source terms become zero from a suitable combination of $\Psi$ and $V$ properties. Apparently, this would require a fine tuning in $\Psi$ and $V$ and the emerging conservation law would not necessarily correspond to a symmetry of the potentials. Nevertheless, this case will not be considered here due to the lack of a specific example leading to a non-trivial generalized current.  

\begin{figure}[h!]
\centering \includegraphics[width=1\columnwidth]{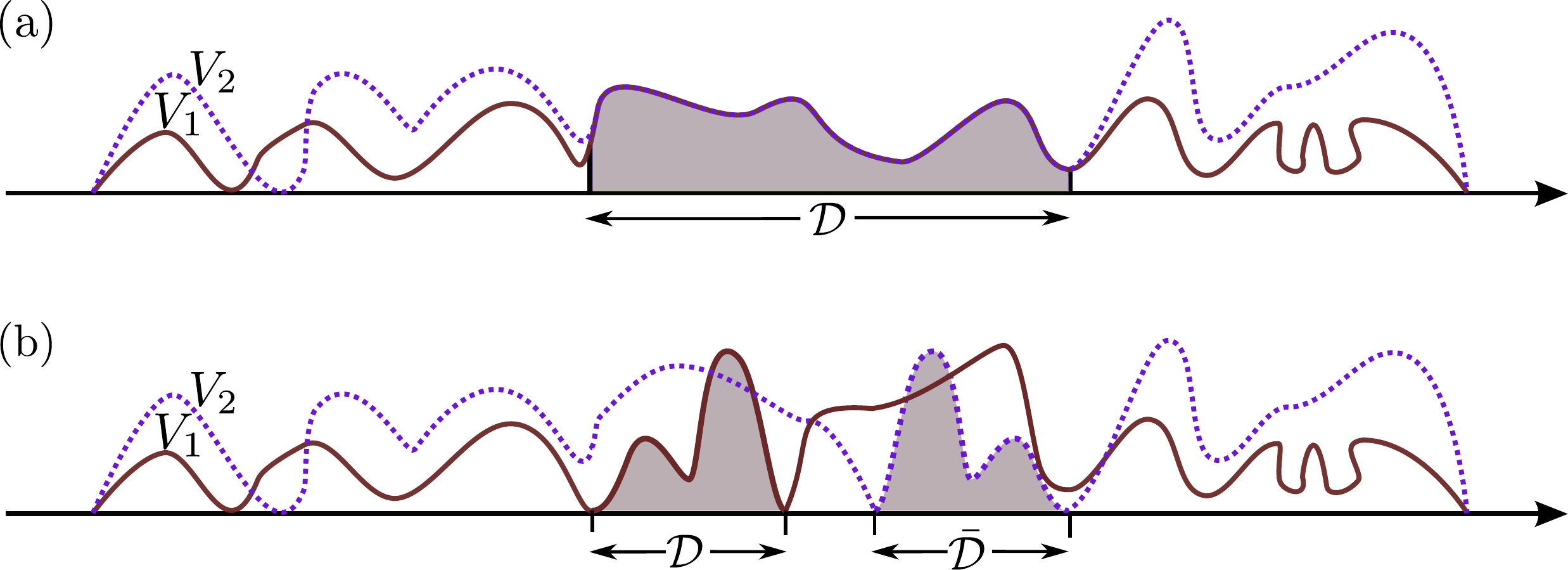}
\caption{(a) Schematic of two different potential landscapes $V_1$ and $V_2$ of two distinct setups which have a common profile indicated by the region $\mathcal{D}$. (b) Schematic of two different potential landscapes $V_1$ and $V_2$ of two distinct setups. The domain $\mathcal{D}$ of $V_1$ is linked to the domain $\bar{\mathcal{D}}$ of $V_2$ via an inversion symmetry transform.}\label{fig1}
\end{figure}

The most straightforward way to set the source terms to zero is by setting the coefficients $C_b=0$. Observing the explicit form of the coefficients as shown in the Appendix \ref{Vapp} and for a diagonal matrix $V$, as in our case, this condition can be realized if in a spatial domain $\mathcal{D}$ holds, 
\begin{equation}
\label{eq:19}
V_i(\vec x)=V_j(\vec x) ~~~;~~~ i\neq j
\end{equation}
where $i,j=1,2,\cdots,N$, enumerate the different components of $\Psi$.  
Obviously, this means that in every region $\mathcal{D}$ where Eq.~\eqref{eq:19} holds, a \textit{locally} conserved  4-current emerges. Such a case, where two different potential landscapes of two distinct problems, have a common profile within a region $\mathcal{D}$, is schematically shown in Fig.~\ref{fig1} (a).
Note, that since the potential term $V_i$, which appears in the $i-$th Dirac equation, is a matrix $ V_i=V_i\cdot \mathds{1}_4$, the coefficients $C_b$ will also be matrices $C_b=C_b\cdot\mathds{1}_4$. This, in turn, leads to the commutation property
\begin{equation}
\label{eq:20}
\bar\Psi C_bf_{abc}T_c \Psi = C_bf_{abc}\bar\Psi T_c\Psi,
\end{equation}
which can be generalized for the case where the potential terms and the coefficients $C_b$ are not proportional to the unit matrix but to $\gamma^0$, i.e. $V_i=V_i\cdot \gamma^0$ and $C_b=C_b\cdot \gamma^0$. Then, Eq.~(\ref{eq:20}) becomes
\begin{equation}
\label{eq:21} 
\bar\Psi C_bf_{abc}T_c \Psi = C_bf_{abc}\Psi^\dagger T_c\Psi.
\end{equation}
	
A similar procedure can be developed in the presence of gauge fields. Using an $SU(N)$ \textit{gauge} transformation of $\Psi$, the Lagrangian density $\mathcal{L}_0$ becomes $\mathcal{L}_0= i\slashed{D}\otimes\mathds{1}_N$ with $\slashed{D}= \gamma^{\mu} D_{\mu}$ and $D_{\mu}=\partial_{\mu} + i A^a_{\mu} T_a$ the covariant derivative containing the gauge field $A^a_{\mu}$. Following the process described above, we find that the corresponding GCE is, 
\begin{equation}
\label{eq:22} \partial_\mu(\bar\Psi\gamma^\mu T_a \Psi-R^{\mu\nu d}f_{ab}^dA_\nu^b)=f_{abc}{C_b}\bar\Psi T_c\Psi.
\end{equation}
The detailed derivation of Eq.~(\ref{eq:22}) can be found in the Appendix~\ref{gaugegce}.

\section{Generalized  continuity equation for two Dirac Systems}\label{sec_III}

To illustrate the properties of the Dirac GCE in a more transparent way, we will focus on the case of two distinct systems each consisting of one Dirac Fermion. This composite system is described by the super-Lagrangian $\mathcal{L}=\mathcal{L}_1+\mathcal{L}_2$, 
\begin{equation}
\label{eq:23}
\mathcal{L}= \left[\begin{array}{cc} \bar\psi_1 & \bar\psi_2 \end{array}\right] \left[\begin{array}{cc} i\slashed{\partial}-V_1&0 \\ 0 & i\slashed{\partial}-V_2 \end{array}\right] \left[\begin{array}{c} \psi_1 \\ \psi_2 \end{array}\right].
\end{equation} 
Recalling that the representation of the $SU(2)$ generators are $T_a=\sigma^a/2$, where $\sigma^a$ are the Pauli matrices, and that the 3D Levi-Civita symbol $\epsilon_{abc}$ corresponds to the structure constants of the group ($f_{abc}=\epsilon_{abc}$), we find that, for a global transformation, the emerging GCEs, which hold for any pair of arbitrary states $\psi_1,\psi_2$, according to Eq.~(\ref{eq:18}) are,
\begin{align}
\partial_\mu(\bar\psi_{1}\gamma^\mu\psi_{2})&=i(V_{1}-V_{2})\bar\psi_{1}\psi_{2}\label{eq:24} \\
\partial_\mu(\bar\psi_{2}\gamma^\mu\psi_{1})&=i(V_{2}-V_{1})\bar\psi_{2}\psi_{1}\label{eq:25} \\
\partial_\mu(\bar\psi_{1}\gamma^\mu\psi_{1})&=\partial_\mu(\bar\psi_{2}\gamma^\mu\psi_{2})\label{eq:26},
\end{align}
where the $C_b$ coefficients are given by Eq.~\eqref{eq:9}. Obviously Eq.~\eqref{eq:26} is trivial and Eqs.~\eqref{eq:24} and \eqref{eq:25} are Hermitian conjugates. Hence, for the rest of our analysis, and without loss of generality, we consider only Eq.~\eqref{eq:24}, aiming to determine the conditions which eliminate the right-hand-side (r.h.s.) of Eq.~\eqref{eq:24}.



\subsection{Global conservation laws}

We first consider the  global conservation case i.e $V_1(\vec x)=V_2(\vec x)~~\forall \vec{x}$, where  the source term in Eq.~\eqref{eq:24} vanishes in the entire space. An interesting scenario occurs when Eq.~\eqref{eq:24} involves eigenstates of the same Dirac  equation which -in the absence of degeneracy- differ in energy. Note, that this holds always for 1-D systems. In higher dimensions, different eigenstates which share the same energy eigenvalue constitute also a possible realization of the global conservation scenario.

Assuming 1-D solutions of the form 
 \begin{equation}
 \label{eq:27}
 \psi_i(\vec{x},t)=e^{-iE_i t}\phi_i(\vec{x}) ~~~;~~~i=1,2
 \end{equation}
and inserting Eq.~(\ref{eq:27}) into Eq.~\eqref{eq:24}, we find 
\begin{eqnarray}
\label{eq:28} \partial_{t} [ \bar{\phi}_{1}(x) e^{iE_1 t} \gamma^{0} \phi_{2}(x) e^{-iE_2 t} ] +  \nonumber \\
  e^{i(E_1-E_2) t} \frac{d}{dx} [\bar{ \phi}_{1}(x) \gamma^{1} \phi_{2}(x)]=0,
\end{eqnarray}
The quantity $J_{12}(x)=\bar{\phi}_{1}(x) \gamma^{1} \phi_{2}(x)$ is identified as the spatial part of the conserved current. The spatial integral, over $x$, of the time derivative in the l.h.s. of Eq.~(\ref{eq:28}) leads to the condition,
\begin{equation}
\label{eq:29} 
\int_{x_1}^{x_2} \bar{\phi}_{1}(x)  \gamma^{0} \phi_{2}(x)  dx =i\small \frac{J_{12}(x_2)-J_{12}(x_1)}{(E_1-E_2)}.
\end{equation}
If $x_1 \rightarrow -\infty$ and $x_2 \rightarrow =+\infty$, we recover the global generalized charge  $Q=\int_{-\infty}^{+\infty} \bar{\phi}_{1}(x)  \gamma^{0} \phi_{2}(x) dx$. This, in turn, can be expressed via currents $J_{12}$ at $\pm \infty$, according to Eq.~(\ref{eq:29}).



\subsection{Local conservation laws} 

Equation~\eqref{eq:24} also supports the scenario of local conservation laws. We assume that the equality $V_1(\vec{x})=V_2(\vec{x})$ holds only in a finite spatial domain $\mathcal{D}$, as shown in Fig.~\ref{fig1} (a) and that there is, at least, one common energy state $E_1=E_2=E$ for the two distinct systems involved in the Lagrangian \eqref{eq:23}. Even though, the two eigenstates share the same eigenvalue, this is not a degeneracy case. These eigenstates correspond to different Dirac equations, e.g. in Fig.~\ref{fig1} (a)  there are $\vec{x}$-regions in which $V_1(\vec{x}) \neq V_2(\vec{x})$.  Otherwise, this would be a case of global conservation law. Note that restricting our analysis to systems which a common energy value simplifies significantly the GCE and only spatial variations remain. This condition can be always satisfied for two different scattering systems. For bound systems, on the other hand, it can be achieved only with appropriate fine tuning. 


Due to the condition of Eq.~\eqref{eq:19} which holds for $\vec{x} \in \mathcal{D}$ [see Fig.~\ref{fig1} (a)], we get
\begin{equation}
\label{eq:30}
\vec{\nabla}(\bar\psi_{1}(\vec x)\vec\gamma\psi_{2}(\vec x))=0  ~~~,~~~\vec x~\in~ \mathcal{D}.
\end{equation}
This, in turn, leads to the divergence-free current,
\begin{equation}
\label{eq:31}
\vec J_{12}(\vec x)=\bar\psi_{1}(\vec x)\vec\gamma\psi_{2}(\vec x) ~~~,~~~\vec x~ \in~ \mathcal{D}.
\end{equation}
The subscripts $1,~2$ denote the solutions of the Dirac equations which involve the potentials $V_{1}$ and $V_{2}$, respectively. 
The  vanishing divergence holds also for the Hermitian conjugate $J_{12}^\dagger$ of the current $J_{12}$. 
It should be stressed here that this current is a completely different quantity from the usual probability current. In fact, its existence is not attributed to the presence of a particular source in region $\mathcal{D}$ [see Fig.~\ref{fig1} (a)], but it is based on the fact that outside region $\mathcal{D}$ the potential landscapes in the two problems differ from each other. 
Restricting our analysis to 1-D, Eq.~(\ref{eq:30}) becomes,
 \begin{equation}
 \label{eq:32}
\partial_x(\bar\psi_{1}(x)\gamma^x\psi_{2}(x))=0,
\end{equation}
yielding a spatially constant current $\forall x \in \mathcal{D}$
\begin{equation}
\label{eq:33}
J_{12}=\bar\psi_{1}(x)\sigma_x\psi_{2}(x)=\text{const.} ~~~,~~~ x~ \in~ \mathcal{D},
\end{equation}
where we have employed the usual convention for the $\gamma$-matrices in 1-D,
\begin{equation}
\label{eq:34}
\gamma^0=\sigma_z ~~~;~~~ \gamma^x=\sigma_x
\end{equation} 
The constant current in $\mathcal{D}$ acts as a correlator between the solutions of two different 1-D Dirac systems, based on the equality of the potentials $V_1(x)$ and $V_2(x)$ in this region. Obviously, if the condition $V_1(x)=V_2(x)$ holds for more than one regions $\mathcal{D}$, i.e in the regions $\mathcal{D}_j$, with $j=1,2,\cdots,M$, then the respective currents $J_{j,12}=c_j$ will be constant in every region $\mathcal{D}_{j}$ while, in general, $c_i \neq c_j$ for $i \neq j$.  

\begin{figure}[h!]
\centering \includegraphics[width=1\columnwidth]{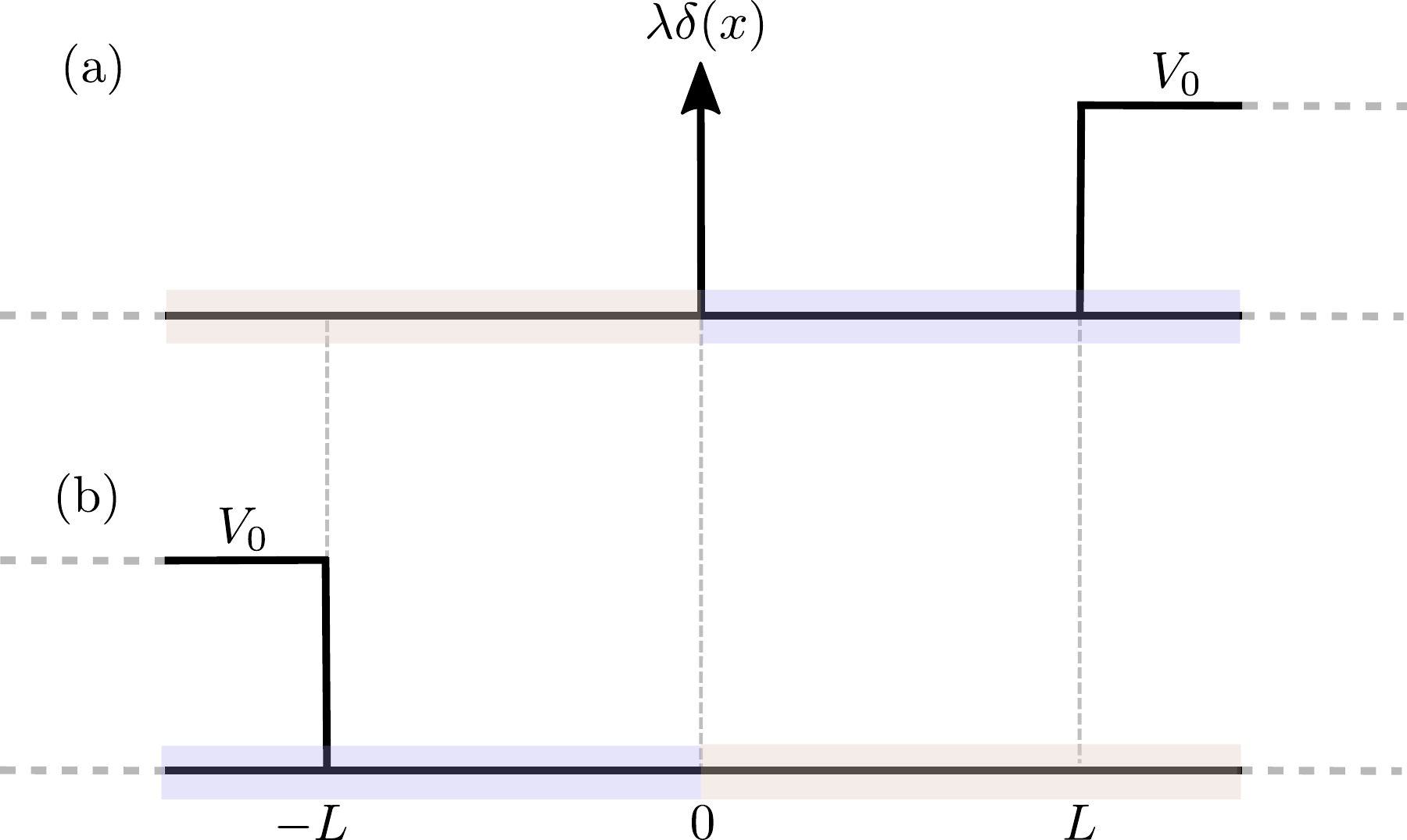}
\caption{Potential landscapes which are linked via an inversion symmetry transformation with the exception of $x=0$ where a $\delta-$barrier is located. The color coding indicates the areas of the wavefunctions which take part in the construction of the respective generalized currents.}\label{fig2}
\end{figure}

Finally, we discuss an interesting sub-case which also belongs to the class of local conservation laws. It is based on a relation of the form,
\begin{equation}
\label{eq:35}
V_1(x)=V_2[F(x)],
\end{equation}
holding within a finite domain $\mathcal{D}$. In Eq.~\eqref{eq:35} $F(x)$ is a linear transformation under which the Laplacian operator is invariant. Notice that since the Dirac field is not a scalar, one has to take into account the spinor transformation under the specific geometric transform (leaving the Laplacian invariant) in the super-Lagrangian. To explain this in more detail let us restrict to the one spatial dimension. In 1-D, $F(x)$ describes either an inversion $\Pi$ with respect to a point $a$ or a translation $T$ by $L$, as shown in Fig.~\ref{fig1} (b). Treating such transforms requires attention. For instance,  the transformed spinor, in the inversion case, is given by $\psi'(x')=\Pi \psi[F(x)] = \sigma_z \psi(-x)$.  The symmetry transform $F(x)$ can be formally written as,
\begin{equation}
\label{eq:36}
F: ~ x \mapsto \overline{x} = F(x) = \sigma x + \rho,
\end{equation}
where the parameters $\sigma$ and $\rho$ are as follows:
\begin{eqnarray}
\sigma=-1,~ &\rho=2 a ~ \Rightarrow F = \Pi &~( \textrm{Parity}) \label{eq:37}\\ 
\sigma=+1,~ &\rho=L ~ \Rightarrow  F = T  &~(\textrm{Translation})
\label{eq:38}
\end{eqnarray}
The symmetry condition \eqref{eq:35} leads, in turn, to a current which is spatially conserved in $\mathcal{D}$ and is expressed as,
\begin{equation}
\label{cons_current}
J_{F,21}=\bar\psi_{1}(x)\sigma_x\psi_{2}(\sigma x +\rho)=\text{const.}
\end{equation}
or
\begin{equation}
\label{eq:39}
J_{F,21}=\bar\psi_{1}(x)\sigma_x\psi'_{2}(x')=\text{const.}
\end{equation}
Note, that the conservation of the generalized current $J_{F,21}$ in the spatial domain $\mathcal{D}$ and its symmetry-transformed counterpart $\bar{\mathcal{D}}$ involves the original $\psi_1(x)$ solution which corresponds to the potential landscape $V_1$ and the transformed solution $\psi'_2(x')$ which corresponds to the potential landscape $V_2$.
We will demonstrate the construction of such a current using the setup illustrated in Figure \ref{fig2}. 
The landscapes $(a)$ and $(b)$ are symmetric under an inversion symmetry transformation, with the exception of the point $x=0$ where the $\delta$-barrier $\lambda \delta(x)$ ($\lambda > 0$) is located in $(a)$. The $\delta$-barrier separates the $x$-space into two distinct domains, namely $D_-=(-\infty,0)$ and $D_+=(0,+\infty)$. The generalized current acquires a different value in each of these two domains. The coloring indicates the wavefunction domains participating in the generalized current in each case. Based on Eq.~\eqref{eq:39} we have : 
\begin{equation}
J_{12}=\bar\psi_1(x)\sigma_x\sigma_z\psi_2(-x)=
\begin{cases}
c_{-} &,~~ x\in D_-\\
c_{+} &,~~ x\in D_+
\end{cases}
\end{equation}
The constants $c_{\pm}$ can be expressed through the left $\psi_i(0^-)$ and right $\psi_i(0^+)$ ($i=1,~2$) limiting values of the Dirac field at $x=0$, as, 
\begin{equation}
J_{12}=
\begin{cases}
\bar\psi_1(0^-)\sigma_x\sigma_z\psi_2(0^+) &,~~ x\in D_-\\
\bar\psi_1(0^+)\sigma_x\sigma_z\psi_2(0^-) &,~~ x\in D_+,
\end{cases}
\end{equation}
Combining the properties of the Pauli matrices, the definition of $\bar\psi(x)$ in 1-D, and the following expressions as given in \cite{1ddirac} :  
\begin{align}
\psi_1(0^+)&=\exp\left({-i\lambda \sigma_y}\right)\psi_1(0^-) \\
\psi_2(0^+)&=\psi_2(0^-),
\end{align}
we find that the constant generalized current can be expressed at $x=0$ as, 
\begin{equation}
J_{12}=
\begin{cases}
-\psi^\dagger_1(0^-)\sigma_x\psi_2(0^-) &,~~ x\in D_-\\
-\psi^\dagger_1(0^-)\exp\left({i\lambda \sigma_y}\right)\sigma_x\psi_2(0^-) &,~~ x\in D_+
\end{cases}
\end{equation}

Since $\exp\left(i\lambda\sigma_y\right)$ represents  
a rotation matrix:
\begin{equation}
\exp\left(i\lambda\sigma_y\right)=\left(\begin{array}{cc} \cos\lambda & \sin\lambda \\ -\sin\lambda & \cos\lambda \end{array}\right)
\end{equation}
we observe that the current $J_{12}$ in $D_+$ differs from $J_{12}$ in $D_-$ in containing the spinor $\sigma_x \psi_2(0^-)$ with a phase shift induced by the $\delta$-barrier. The phase shift angle is given by $\lambda$. Obviously, when $\lambda=0$ there is a single value for the constant $J_{12}$ in the entire $x$-space. 

\section{$SU(N)$ complete set of GCEs for Schr\"odinger systems}\label{sec_IV}

In this section we return to the Schr\"odinger framework and we generalize the results obtained in \cite{Diakonos2019} for non-relativistic quantum systems. To this end, we use a complete set of $SU(N)$ transformations instead of those belonging to the Cartan sub-algebra, as in \cite{Diakonos2019}. In addition, we include a gauge field in our analysis. The super-Lagrangian in the Schr\"{o}dinger framework is written as,
\begin{equation}
\label{eq:40}
\mathcal{L}=\sum_{i=1}^N\mathcal{L}_i=\Psi^\dagger(H_0-V)\Psi=\mathcal{L}_0+\mathcal{\tilde L}+h.c,
\end{equation}
with
\begin{equation}
\label{eq:41}
\mathcal{L}_i=\Psi_i^\star(\vec{x},t)\left[i\partial_t-\frac{\hat \vec{p}^2}{2m}-V_{i}(\vec{x})\right]\Psi_i(\vec{x},t)~+~h.c.
\end{equation}
and 
\begin{equation}
\label{eq:42}
\Psi=\begin{pmatrix} \Psi_1(\vec{x},t) \\ \Psi_2(\vec{x},t) \\ \vdots \\ \Psi_N(\vec{x},t) \end{pmatrix}.
\end{equation}
In Eqs.~\eqref{eq:41},~\eqref{eq:42}, $\Psi_{i}(\vec{x},t)$ is the wavefunction of the $i$-th system involved in the Schr\"{o}dinger super-Lagrangian. Obviously, 
\begin{equation}
\label{eq:43}
\mathcal{L}_0=\Psi^\dagger H_0 \Psi  ~~~;~~~\mathcal{\tilde L}=-\Psi^\dagger V \Psi
\end{equation}
with
\begin{equation}
\label{eq:44}
H_0=(i\partial_t-\frac{\hat \vec{p}^2}{2m})\otimes\mathds{1}_N
\end{equation}
and
\begin{equation}
\label{eq:45}
V=\text{diag}(V_1,V_2,\hdots,V_N)
\end{equation}
The symmetry transformation of Eq.~\eqref{eq:8} leads directly to the variation relations $\delta \mathcal{L}_0=0,~\delta\mathcal{\tilde L}\neq0$ and as expected, once again $\delta \mathcal{L}=\delta\mathcal{\tilde L}\neq0$.
%
Since $\delta \mathcal{\tilde{L}}=\mathcal{\tilde L}'-\mathcal{\tilde L}$ we find that the Lagrangian density variation is given by
\begin{equation}
\label{eq:46}
\delta\mathcal{L}=- 2\Psi^\dagger(U^\dagger V U - V)\Psi. 
\end{equation}
Substituting Eq.~\eqref{eq:9} we find that,
\begin{equation}
\label{eq:47}
U^\dagger V U - V=-\delta \theta_a C_kf_{ak}^cT_c
\end{equation} 
and consequently
\begin{equation}
\label{eq:48}
\delta \mathcal{L}=2\delta \theta^a C_bf_{abc}\Psi^\dagger T_c\Psi.
\end{equation}


The variation of $\delta\mathcal{L}$ can also be written as 
\begin{align}
\label{eq:49}
\delta\mathcal{L}=\sum_{i=1}^N [ \frac{\partial \mathcal{L}}{\partial \Psi_i}\delta\Psi_i+\frac{\partial \mathcal{L}}{\partial( \partial_\mu\Psi_i)}\delta(\partial_\mu\Psi_i)+  \nonumber \\
\frac{\partial \mathcal{L}}{\partial {\Psi^\star_i}}\delta{\Psi^\star_i}+\frac{\partial \mathcal{L}}{\partial (\partial_\mu{{\Psi_i^\star})}}\delta(\partial_\mu{\Psi^\star_i})],
\end{align}
which after some algebraic calculations leads to,
\begin{equation}
\label{eq:50}
\delta\mathcal{L}=[\partial_t(\Psi^\dagger T_a\Psi)+\frac{i}{m}\vec{\nabla}[\vec{\nabla}(\Psi^\dagger)T_a\Psi]+h.c]\delta \theta^a.
\end{equation}
Combining Eq.~(\ref{eq:48}) and Eq.~(\ref{eq:50}) we finally get
\begin{align}
\label{eq:51}
\partial_t(\Psi^\dagger T_a\Psi)+\frac{i}{2m}\vec\nabla(\vec\nabla\Psi^\dagger T_a\Psi-\Psi^\dagger T_a\vec\nabla\Psi) =  \nonumber \\ f_{abc}C_b\Psi^\dagger T_c\Psi
\end{align} 
which generalizes the GCE found in \cite{Diakonos2019}. 

Note, that the source terms both in the Schr\"odinger and the Dirac GCEs have the same form [see Eq.~\eqref{eq:20} and Eq.~\eqref{eq:51}]. The conservation of the corresponding generalized currents implies the same conditions both for the relativistic and the non-relativistic case. On the contrary, the form of the conserved current differs between these two cases, i.e. in the non-relativistic GCE the current contains derivatives of the wave field while in the relativistic version there are no derivatives involved. This is due to the difference in the form of the kinetic term in either case.

\section{Summary \& Conclusions}  \label{sec_V}

In this work we develop a systematic approach to derive generalized 4-currents built from fermionic fields which describe solutions of Dirac problems involving different external potential terms. This is achieved through the construction of a super-Lagrangian which allows the synthesis of a composite system acting as a generator for all involved sub-problems along the lines presented in \cite{Diakonos2019} in the context of Schr\"{o}dinger (non-relativistic) quantum mechanics. The derived currents are related to symmetry transforms of the super-Lagrangian and they obey generalized continuity equations which involve source terms dictated by the specific transform. These source terms occur because the super-Lagrangian is not invariant under this transformation. Following this procedure, we have explored possible scenarios based on conditions on the form of the external potentials leading to the vanishing of the source terms globally or in finite spatial domains. In the second case local conservation laws manifest, relating solutions of Dirac equations which differ in the external potential terms. We have illustrated in a simple example how such a current is constructed in practice. In contrast to the Schr\"{o}dinger case presented in \cite{Diakonos2019} the generalized Dirac currents do not involve derivative terms and therefore they can be considered as a direct mapping between solutions of different Dirac equations. It would be interesting to apply this procedure to specific problems and explore the insight gained by such a mapping. The corresponding local conservation laws may be useful for the control as well as the preparation of fermionic states through suitable modulation of the external potential(s).
Here we have considered exclusively $SU(N)$ transformations of the super-field formed from the solutions of the different Dirac problems. Future perspectives of this work would be the investigation of other symmetry transformations and how new conservation laws could emerge.

\appendix
\section{Analysis of V-matrix to SU(N) Representations of Generators}\label{Vapp}

A general $N\times N$ $V$-matrix is analyzed to the basis of the generators of the $SU(N)$ group using the formula,
 \begin{equation*}
V=\left(\begin{array}{ccc} V_{11} & \cdots & V_{1N} \\ \vdots & \ddots & \vdots \\ V_{N1} & \cdots & V_{NN}  \end{array}\right)=V_0\mathds{1}_N+\sum_{k=1}^{N^2-1} C_kT_k
\end{equation*}
where
\begin{equation*}
V_0=\frac{Tr(V\cdot \mathds{1}_N)}{N} ~~~ ; ~~~ C_k=Tr(V\cdot T_k).
\end{equation*} 
The explicit form of the coefficient $V_0$ is,
\begin{equation}
V_0=\frac{V_{11}+V_{22}+\cdots+V_{NN}}{N}
\end{equation}
For $k=n^2-1$ ($n=2,3,\dots,N$), the coefficients $C_k$ which are related to the diagonal representations of the generators are,
\begin{align*}
C_3&=V_{11}(x)-V_{22}(x) \\
C_8&=\frac{V_{11}(x)+V_{22}(x)-2V_{33}(x)}{\sqrt{3}}\\
\vdots~~&\qquad\qquad~~~\vdots \\
C_{N^2-1}&=\frac{V_{11}(x)+\cdots-(N-1)V_{NN}(x)}{\sqrt{\frac{N(N-1)}{2}}}. 
\end{align*}
Accordingly, the coefficients which are related to the non-diagonal representations of the generators are,
\begin{align*}
C_{k}&=V_{k,k+1}(x)+V_{k+1,k}(x) \\
C_{k+1}&=[V_{k,k+1}(x)-V_{k+1,k}(x)]i \\
\end{align*}
where,
 \begin{align*}
 k&=n^2+2m \\
n&=1,2,\dots,N \\
m&=0,1,\dots,n-1.
 \end{align*}
Employing our analysis for  the full-element $V$-matrix, we derive the GCEs for an $SU(N)$ global infinitesimal transformation for $N$ Dirac problems:
 \begin{align*}
\partial_\mu(\bar\psi_i\gamma^\mu\psi_j)&=i(V_{ii}-V_{jj})\bar\psi_i\psi_j\\&+iV_{ji}(\bar\psi_j\psi_j-\bar\psi_i\psi_i)\\&+iV_{ki}\bar\psi_k\psi_j-iV_{jk}\bar\psi_i\psi_k \\
\partial_\mu(\bar\psi_i\gamma^\mu\psi_i-\bar\psi_j\gamma^\mu\psi_j)&=2iV_{ji}\bar\psi_{j}\psi_{i}-2iV_{ij}\bar\psi_{i}\psi_{j}\\&+iV_{jk}\bar\psi_{j}\psi_{k}-iV_{kj}\bar\psi_{k}\psi_{j}\\&+iV_{ki}\bar\psi_{k}\psi_{i}-iV_{ik}\bar\psi_{i}\psi_{k}
\end{align*}
where 
\begin{align*}
i,j,k=1,2,\dots,N \\
i\neq j\neq k
\end{align*}
and $k$ sum over all the remaining indices. 
The non-diagonal terms express a direct connection between the $i-$th Dirac problem with the $N-1$ remaining problems. 

\section{Gauge Transformation and GCE}\label{gaugegce}

We present the basic steps towards a GCE which is induced by a gauge transformation, for $N$ Dirac problems.
A gauge symmetric Lagrangian under a $U=e^{-i\theta^a(x)T_a}$ transformation is,
\begin{equation*}
{\mathcal{L}=\bar{\Psi}(i\slashed{D}-m)\Psi-\frac{1}{4}R_{\mu\nu}^aR^{\mu\nu a}}
\end{equation*}
where
\begin{align*}
D_\mu&=\partial_\mu+iA^a_\mu T_a\\
R_{\mu\nu}^a&=\partial_\mu A_\nu^a -\partial_\nu A_\mu^a-f_{bc}^aA_\mu^b A_\nu^c
\end{align*} 
and $A_\mu^a$ are scalar fields.
Demanding $D_\mu'\Psi'=UD_\mu\Psi$, this Lagrangian becomes invariant under the corresponding gauge transformation. 
The Lagrangian of these different problems is,
\begin{equation*}
{\mathcal{L}=\bar{\Psi}(i\slashed{\partial}-\slashed{A^a}T_a-V)\Psi-\frac{1}{4}R_{\mu\nu}^aR^{\mu\nu a}},
\end{equation*}
where $V$ is a diagonal matrix. In that case the Lagrangian becomes,
\begin{align*}
\mathcal{L}=\bar\Psi(i\slashed{D}-V)\Psi-\frac{1}{4}R_{\mu\nu}^aR^{\mu\nu a}=\mathcal{L}_0+\tilde{\mathcal{L}}
\end{align*}
with
\begin{align*}
	\mathcal{L}_0&=\bar\Psi i\slashed{D}\Psi-\frac{1}{4}R_{\mu\nu}^aR^{\mu\nu a} \\
\tilde{\mathcal{L}}&=-\bar\Psi V\Psi
\end{align*}
Following the same procedure we find that
\begin{align*}
	{\delta\tilde{\mathcal{L}}=\bar\Psi\theta^a(x){C_b}f_{ab}^cT_c\Psi},
\end{align*}
and
\begin{equation*}
 \delta \mathcal{L}=\partial_\mu(\bar{\Psi}\gamma^\mu T_a\Psi+R^{\mu\nu c}f_{ba}^cA_\nu^b)\theta^a(x).
 \end{equation*}
Finally, we find that the GCE generated by a gauge $SU(N)$ transformation is,
\begin{align*}
	\partial_\mu(\bar\Psi\gamma^\mu T_a \Psi-R^{\mu\nu d}f_{ab}^dA_\nu^b)=\bar\Psi\ C_bf_{ab}^cT_c\Psi.
	\end{align*}
This GCE is similar to the GCEs for global transformations, having only an extra term in the 4-gradient, as expected.

{}

\onecolumngrid
\newpage
\begin{table}[h]
\caption*{Summary of GCEs for Schr\"{o}dinger and Dirac Problems}
\centering
\resizebox{\textwidth}{10cm}{
\begin{tabular} {| C{2cm} | C{8.5cm} | C{8.5cm} |}
\hline
 & Schr\"{o}dinger & Dirac \\
 \hline
\makecell{SU(N) \\3-D
}
&{\begin{equation*}
\partial_t(\Psi^\dagger T_a\Psi)+\frac{i}{2m}\vec\nabla(\vec\nabla\Psi^\dagger T_a\Psi-\Psi^\dagger T_a\vec\nabla\Psi)=C_bf_{abc}\Psi^\dagger T_c\Psi
\end{equation*}}
 &{	\begin{equation*}
	\partial_\mu(\bar{\Psi}\gamma^\mu\otimes T_a\Psi )=f_{abc}\bar\Psi C_b\otimes T_c\Psi
	\end{equation*}}
\\
Common Energy Condition
&{\begin{equation*}
\frac{i}{2m}\vec\nabla(\vec\nabla\Psi^\dagger T_a\Psi-\Psi^\dagger T_a\vec\nabla\Psi)=C_bf_{abc}\Psi^\dagger T_c\Psi
\end{equation*}}
 &{	\begin{equation*}
	\vec\nabla(\bar{\Psi}\vec\gamma \otimes T_a\Psi )=f_{abc}\bar\Psi C_b\otimes T_c\Psi
	\end{equation*}}
\\
Vanishing Source Terms Condition 
&{\begin{equation*}
\frac{i}{2m}\vec\nabla(\vec\nabla\Psi^\dagger T_a\Psi-\Psi^\dagger T_a\vec\nabla\Psi)=0
\end{equation*}}
 &{	\begin{equation*}
	\vec\nabla(\bar{\Psi}\vec\gamma \otimes T_a\Psi )=0
	\end{equation*}}
\\
Divergence Free Currents : 
&\begin{equation*} 
\vec {J}_{12}^a(\vec x)=\frac{i}{2m}\left[\vec\nabla\Psi^\dagger(\vec x) T_a\Psi(\vec x)-\Psi(\vec x)^\dagger T_a\vec\nabla\Psi(\vec x)\right] \end{equation*} & 
\begin{equation*} 
\vec{J}_{12}^a(\vec x)=\bar\Psi(\vec x)\vec \gamma\otimes T_a\Psi(\vec x)
\end{equation*}
\\
\hline
\makecell{
SU(N) \\ 1-D}
&{\begin{equation*}
\partial_t(\Psi^\dagger T_a\Psi)+\frac{i}{2m}\partial_x(\partial_x\Psi^\dagger T_a\Psi-\Psi^\dagger T_a\partial_x\Psi)=C_bf_{abc}\Psi^\dagger T_c\Psi
\end{equation*}}
 &{	\begin{equation*}
	\partial_t(\Psi^\dagger T_a\Psi )+\partial_x(\bar\Psi\gamma^xT_a\Psi)=C_bf_{abc}\bar\Psi T_c\Psi
	\end{equation*}}
	\\
Common Energy Condition
&{\begin{equation*}
\frac{i}{2m}\partial_x(\partial_x\Psi^\dagger T_a\Psi-\Psi^\dagger T_a\partial_x\Psi)=C_bf_{abc}\Psi^\dagger T_c\Psi
\end{equation*}}
 &{	\begin{equation*}
\partial_x(\bar\Psi\gamma^xT_a\Psi)=C_bf_{abc}\bar\Psi T_c\Psi
	\end{equation*}} 
	\\
Vanishing Source Terms Condition   
	&{\begin{equation*}
\frac{i}{2m}\partial_x(\partial_x\Psi^\dagger(x) T_a\Psi(x)-\Psi^\dagger(x)T_a\partial_x\Psi(x))=0
\end{equation*}}
 &{	\begin{equation*}
\partial_x(\bar\Psi(x)\gamma^x\otimes T_a\Psi(x))=0 
   \end{equation*}}
\\
Conserved Currents
& \begin{equation*} 
J^a_{12}=\frac{i}{2m}\left[\partial_x\Psi^\dagger(x) T_a\Psi(x)-\Psi(x)^\dagger T_a\partial_x\Psi\right]=\text{const.} \end{equation*} & 
\begin{equation*} 
J^a_{12}=\bar\Psi(x)\gamma^x \otimes T_a\Psi(x)=\text{const.}
\end{equation*} \\
\hline && \\
\makecell{
SU(2) \\ 1-D}
&{\begin{align*}
&\partial_t(\Psi_1^\star \Psi_2)+\frac{i}{2m}\partial_x(\partial_x\Psi_1^\star\Psi_2-\Psi_1^\star\partial_x\Psi_2)=i(V_1-V_2)\Psi_1^\star\Psi_2\\
&\partial_t(\Psi_2^\star \Psi_1)+\frac{i}{2m}\partial_x(\partial_x\Psi_2^\star\Psi_1-\Psi_2^\star\partial_x\Psi_1)=i(V_2-V_1)\Psi_2^\star\Psi_1 \\
&\partial_t(\Psi_1^\star\Psi_1)+\frac{i}{2m}\partial_x(\partial_x\Psi_1^\star\Psi_1-\Psi_1^\star\partial_x\Psi_1)=\\&\partial_t(\Psi_2^\star\Psi_2)+\frac{i}{2m}\partial_x(\partial_x\Psi_2^\star\Psi_2-\Psi_2^\star\partial_x\Psi_2)
\end{align*}}
&{ \begin{align*}
\partial_\mu(\bar\psi_1\gamma^\mu\psi_2)&=i (V_1-V_2)\bar\psi_1\psi_2 \\
\partial_\mu(\bar\psi_2\gamma^\mu\psi_1)&= i(V_2-V_1)\bar\psi_2\psi_1\\
\partial_\mu(\bar\psi_1\gamma^\mu\psi_1)&=\partial_\mu(\bar\psi_2\gamma^\mu\psi_2)
\end{align*}}
\\
Common Energy Condition
&{\begin{align*}
&\frac{i}{2m}\partial_x(\partial_x\Psi_1^\star\Psi_2-\Psi_1^\star\partial_x\Psi_2)=i(V_1-V_2)\Psi_1^\star\Psi_2
\end{align*}}
&{ \begin{align*}
\partial_x(\bar\psi_1\gamma^x\psi_2)&=i (V_1-V_2)\bar\psi_1\psi_2
\end{align*}}
\\
Condition for $V_1=V_2$
&{\begin{align*}
&\frac{i}{2m}\partial_x(\partial_x\Psi_1^\star\Psi_2-\Psi_1^\star\partial_x\Psi_2)=0
\end{align*}}
&{ \begin{align*}
\partial_x(\bar\psi_1\gamma^x\psi_2)&=0
\end{align*}}
\\
Conserved Current
&{\begin{align*}
&J_{12}=\frac{i}{2m}\left[\partial_x\Psi_1^\star\Psi_2-\Psi_1^\star\partial_x\Psi_2\right]=\text{const.}
\end{align*}}
&{ \begin{align*}
J_{12}=\bar\psi_1\gamma^x\psi_2&=\text{const.}
\end{align*}}
\\
\hline
 \end{tabular}}
 \end{table}

 \end{document}